\documentclass[twocolumn,showpacs,superscriptaddress]{revtex4}

\usepackage{amsmath,amssymb}
\usepackage{graphicx}
\usepackage{dcolumn}
\usepackage{bm}

\begin{document}

\title{Shaping coherent excitation of atoms and molecules \\
by a train of ultrashort laser pulses}

\author{A. Gogyan}
\email{anahit.gogyan@u-bourgogne.fr} \affiliation{Institute for Physical
Research, Armenian National Academy of Sciences, Ashtarak-2, 0203,
Armenia} \affiliation{Laboratoire Interdisciplinaire Carnot de
Bourgogne, UMR CNRS 5209, BP 47870, 21078 Dijon, France}
\author{S. Gu\'erin}
\affiliation{Laboratoire Interdisciplinaire Carnot de Bourgogne,
UMR CNRS 5209, BP 47870, 21078 Dijon, France}
\author{Yu. Malakyan}
\affiliation{Institute for Physical Research, Armenian National
Academy of Sciences, Ashtarak-2, 0203, Armenia}
\affiliation{Centre of Strong Field Physics, Yerevan State
University, 1 A. Manukian St., Yerevan 0025, Armenia}

\date{\today}

\begin{abstract}
We propose a mechanism to produce a superposition of atomic and
molecular states by a train of ultrashort laser pulses combined
with weak control fields. By adjusting the repetition rate of the
pump pulses and the intensity of the coupling laser, one can
suppress a transition, while simultaneously enhancing the desired
transitions. As an example various superpositions of states of the
$K_2$ molecule are shown.
\end{abstract}

 \maketitle

\section{\protect\normalsize INTRODUCTION}

Population transfer to a desired coherent superposition of atomic
and molecular states (i.e. a wavepacket) has been a major goal
during the last three decades and continues to be a challenge for
instance for implementation of chemical and biological processes
\cite{rice, rabitz,wenach}, for fast quantum information
processing \cite{lukin,saffman,jakch,tesch} and for nonlinear
optics \cite{Jain}.

Besides extensions of $\pi$-pulse techniques
\cite{Allen,Shore,Holthaus,lifting} and of brute-force optimal
control \cite{Glaser}, mechanisms to produce superpositions of two
states in atoms based on adiabatic passage (in nanosecond regime)
have been proposed \cite{Yatsenko,lifting,Sangouard} and
demonstrated \cite{Rickes,Sautenkov,Oberst}. Extending such
techniques to an ultrafast regime and for molecular systems is of
particular interest. Another ultrafast spectroscopic technique is Impulsive Stimulated Raman Scattering (ISRS) that provides vibrational structural information with high temporal and spectral resolution \cite{Yan, Nazarkin, Wittmann}. ISRS excitation of coherent phonons, molecular vibrations, and other excitations (including rotational, electronic, and spin) plays important roles in femtosecond pulse interactions with molecules, crystals, glasses (including optical fibers), semiconductors, and metals \cite{Yan}. This is an effective approach  to determine the dynamics of vibrational molecular  motion \cite{Bartels}.

Recent progress has allowed the development
of mode-locked laser systems producing mutually phase-coherent
ultrashort laser pulses of high intensity with arbitrary
controllable amplitudes, of stable frequency and of adjustable
delay time (see for instance \cite{Hansch,Ye}). Theoretical
\cite{Vitanov,FelintoOC2003,Jakubetz} and experimental
\cite{FelintoExp,YeScience} analysis in a few level systems have
shown that a resonant $\pi$-pulse (or generalized $\pi$-pulse
\cite{Holthaus,lifting}) can be split into trains of fractional
$\pi$-pulses and can lead to the accumulation of population in a
target state for appropriate delays. The main point is that weak
pulses can then be used preventing detrimental destructive effects
such as ionization. For more complicated systems, populating some
chosen states among a set of levels all within the broad
ultrashort pulse spectrum is a major issue. However one can
exploit one of the main properties of the associated frequency
comb, that is its extremely small resolution, given by the width
of the comb's teeth in the frequency domain, much better than the
one determined by the Fourier transform of a single pulse in the
train. High degree of population transfer to a single vibrational
state of an electronic excited state has been indeed numerically
shown by a train of fs laser pulses by choosing the pulse
repetition period as noninteger multiple of the vibrational period
\cite{araujo}. Recently a so-called piecewise adiabatic passage
method, based on the combination of adiabatic passage, trains of
pulses, and pulse-shaping techniques, has been proposed
\cite{shapiro,shapiro2}.

In this paper, we propose an alternate robust and efficient method
for population transfer to a desired superposition in multilevel
systems using a train of pulses combined with weak controlled
lasers. We derive analytical formulas in the impulsive and
perturbative regimes for the ultrashort pump pulses.
% We explain in
%particular analytically the numerical result obtained in
%\cite{araujo} concerning the selective excitation of a single
%state.

We consider a system of level configuration shown in Fig. 1
interacting with a train of ultrashort femtosecond laser pulses,
whose spectrum is wide enough to overlap all the upper states,
while a narrow-band weak laser couples, for example, the upper
level 1 with an auxiliary state 4. We consider three upper levels
for simplicity, but the proposed mechanism can be directly
extended to any number of upper-lying levels. By adjusting the
repetition rate of the pump pulses with respect to the Rabi
frequency $\Omega_c$ of the coupling field, this scheme enables
one to cancel out the strong transition $0 \rightarrow 1$ from the
pump field, while enhancing the transitions $0 \rightarrow i,
i=2,3$. To give an insight into the proposed mechanism, let us
consider the interaction of the system with two consecutive
identical pump pulses in resonance on the transition $0
\rightarrow 1$, with a time delay $\tau_d$, which is larger as
compared to the pulse duration $T$. In the low intensity regime,
the atomic state amplitudes $C_{1,2,3}$ after the first pump pulse
are: $C_{j} \sim \theta_{j}  = \int \Omega_{j}(t)e^{i\Delta_j}dt
\ll 1, \; j=1,2,3$,  with $\Omega_{j}$ the pump Rabi frequencies
corresponding to the respective $0 \rightarrow j$ transitions [see
Eqs. \eqref{Field} and \eqref{Rabi} for the definition of the
fields and the Rabi in this paper]. At the end of the second pulse
they take the forms: $C_1\sim\theta_1+ \theta_1\cos(\Omega_c
\tau_d)$ and $C_{2,3}\sim 2\theta_2$ showing that when the delay
time $\tau_d$ is such that
\begin{equation}\label{1}
\Omega_c\tau_d=\pi(1+2k),
\end{equation}
with $k$ an integer, the population on the level 1 vanishes, while
it increases four times on states 2 and 3: The excitation
amplitudes of the two pump pulses add coherently for an
appropriate delay \cite{Vitanov}. Hereafter we assume that the upper-lying
levels are harmonic, such that the condition $\omega_{ij}\tau_d =
2 \pi k_{ij}$ applies, with $k_{ij}$ an integer and
$\omega_{ij}=\omega_j-\omega_i$ the frequency splitting between
the upper levels of energies $\omega_j$. This condition is
essential to accumulate population in the upper states from pulse
to pulse. Therefore, as long as the pulse delay $\tau_d$ remains
well smaller than the atomic decoherence time, the second pulse
allows the selective excitation of a superposition of states 2 and
3. Our method does not suffer from a high sensitivity to
laser-field instabilities. We show below that the efficiency of
the process is preserved even when the condition \eqref{1} is not
well satisfied.

\begin{figure}[b] \rotatebox{0}{\includegraphics* [scale =
0.7]{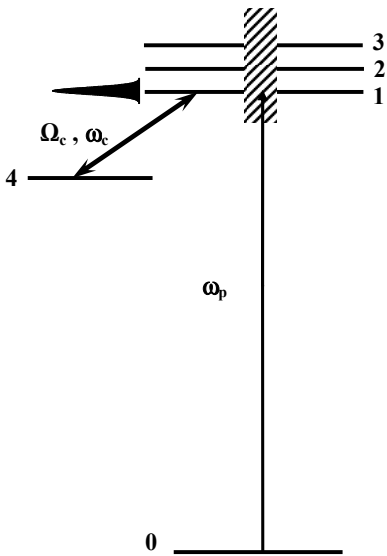}} \caption{Level scheme illustrating the excitation
of three upper states by the ultrashort pump laser. A cw coupling
field drives the auxiliary transition $1\rightarrow 4.$}
\end{figure}
The paper is organized as follows. In the next section we derive
and solve the basic equations for the time evolution of the state
amplitudes in the impulsive and perturbative regimes. In Sec. III
we apply the proposed technique to produce superpositions of
states in an electronic state of the molecule $K_2$. Our
conclusions are summarized in Sec. IV.

\section{Mechanism of selective excitation}
\subsection{The model}
In our scheme (Fig. 1), the upper states 1, 2 and 3 are populated
via a single photon excitation by a train of $m$ identical and
non-overlapping ultrashort laser pulses, whose spectrum is
centered on the resonance with the transition $0 \rightarrow 1$
and is wide enough that upon interacting with each pulse, all the
states 1, 2 and 3 are excited simultaneously: $T^{-1} \sim \Gamma
> \omega_{31}$. Here $\Gamma$ is the spectral width of the laser
fields and $\omega_{31}$ is the frequency splitting of the levels
1 and 3. We assume that the narrow-band coupling field is in exact
resonance with the transition $1 \rightarrow 4$ with the pulse
duration much longer than that of the pump pulses $T$. In what
follows, we neglect the Doppler broadening because it is small as
compared to $\Gamma$.

The  pump $E_{p}(t)$ and coupling $E_{c}( t)$ field amplitudes of
respective carrier frequencies  $\omega_{p}$ and $\omega_{c}$ are
of the form (in complex notation)
\begin{equation}
\label{Field} E_ {p} (t)= \sum_{i=1}^{m} \mathcal E_{i}(t)
e^{i\omega_{p}t}, \quad E_ {c} (t)= \mathcal E_{c}(t)
e^{i\omega_{c}t},
\end{equation}
with the same shape $f(t)$ for all $m$ pump pulses that determines the
time dependence of $\mathcal E_{i}(t)= \mathcal
E_{0}f(t-t_1-(i-1)\tau_d)$, and the delay $\tau_{d}$ between two
consecutive pulses. The interaction of the system with the pump
and coupling fields is determined by their Rabi frequencies at the
corresponding transitions
\begin{equation}\label{Rabi}
\Omega_{p }^{(j)}(t) = \frac{\mu_{j}}{\hbar}\mathcal E_{i}(t),
\quad \Omega_{c }(t) = \frac{\mu_{14}}{\hbar} \mathcal E_{c}(t),
\end{equation}
where $\mu_{ij}$ is the dipole matrix element of the transition $i
\rightarrow j=1,2,3$ and the notation $\mu_j\equiv\mu_{0j}$. We
consider for simplicity a time independent coupling field. Our
results generalize for a pulsed coupling field of much longer
duration than the pump field. In the rotating wave approximation
with respect to the pump field, the Hamiltonian of the system is
given by
\begin{equation} \label{hamilt}
H = - \hbar \sum_{j=1}^3\bigl ( \Omega_{p}^{(j)} \sigma_{j0}-
\Delta_{j} \sigma_{jj}\bigr )-\hbar\Delta_1\sigma_{44}- \hbar  \Omega_c \sigma_{41} + h.c.,
\end{equation}
where $\sigma_{ij}=|i\rangle\langle j|$ are the atomic operators
and $\Delta_{j}=\omega_{j0}-\omega_{p}$ is two-photon detuning of
the pump field from the $0\rightarrow j,\ j=1,2,3$ transition. The
state $|\psi(t)\rangle = \sum_{i} C_i (t) |{i}\rangle$ of the atom
satisfies the Schr\"odinger equation $\dot{C_i}(t) = - \frac{i}
{\hbar} \sum_k \langle i|H|k\rangle C_k(t),$ which leads to the
equations for the atomic state amplitudes
\begin{subequations}
\label{ampl}
\begin{eqnarray} \label{ampl0}
\dot{C}_0(t) &=& i \sum_{j=1,2} \Omega_{p}^{(j)\ast} C_j(t),\\
\label{ampl1} \dot{C}_1(t) &=& -i \Delta_{1} C_1+i
\Omega_{p}^{(1)}
C_0(t)+i \Omega_c C_4,\\
 \label{ampl2} \dot{C}_{2,3}(t) &=& -i
\Delta_{2,3} C_{2,3}+i \Omega_{p}^{(2,3)} C_0(t),\\
\label{ampl3} \dot{C}_4(t) &=&  -i\Delta_1 C_4 +i \Omega_c^{\ast}
C_1
\end{eqnarray}
\end{subequations}
with the initial conditions
\begin{equation}
\label{am1reg1} C_0(-\infty)=1,\quad C_{j\neq 0}(-\infty)=0.
\end{equation}

\subsection{Solution in the impulsive regime}

In the general case, Eqs. (\ref{ampl}) do not provide an analytic
solution. However, in the regime of low intensity of the coupling
field with respect to the pump fields: $\Omega_c \ll \Omega_{p}$
and in the impulsive (or sudden) approximation for the ultrashort pump pulse
by disregarding the detunings $\Delta_jT\ll1$ \cite{UTDST}, one
can determine the solution (see Appendix). Equations (\ref{itC})
show the dependence of the state amplitudes after  the $(n+1)$-st
pulse depending on the amplitudes after the $n$-th pulse
($n=1,2,...$). For the sequence of two pump pulses
%We obtain after the first pump pulse:
%\begin{subequations}
%\begin{eqnarray}
%C_0(t_1^+) &=& \cos \theta\\
%C_j(t_1^+) &=& i \frac{\mu_j}{\mu} \sin \theta,\quad j=1,2,3\\
%C_4(t_1^+)&=&0,
%\end{eqnarray}
%\end{subequations}
%and right before the second pump pulse, i.e. for $t=t_2^-$:
%\begin{subequations}
%\begin{eqnarray}
%C_0(t_2^-) &=& \cos \theta\\
%C_j(t_2^-) &=& i\frac{\mu_j}{\mu}e^{ -i \Delta_{j}\tau_d}  \sin
%\theta,\quad j=2,3,\\
%C_1(t_2^-) &=& i \frac{\mu_1}{\mu}e^{- i \Delta_{1}\tau_d}
%\sin \theta \cos(\Omega_c \tau_d),\\
%C_4(t_2^-) &=& - \frac{\mu_1}{\mu}e^{ -i \Delta_{1} \tau_d}  \sin
%\theta \sin (\Omega_c \tau_d).
%\end{eqnarray}
%\end{subequations}
right after the interaction with the second pump pulse, Eqs.
(\ref{itC}) lead to
%\begin{equation}
%\bar C_2^-=\frac{i}{\mu^2}\sin \theta\Bigl[|\mu_1|^2 e^{- i
%\Delta_{1}\tau_d}  \cos(\Omega_c \tau_d) +\sum_{k=2}^3|\mu_k|^2e^{
%-i \Delta_{k}\tau_d}\Bigr]
%\end{equation}
%and
\begin{widetext}
\begin{subequations}
\begin{eqnarray}
C_0(t_2^+)&=&\cos^2 \theta -\frac{1}{\mu^2} \bigl(\mu_1^2e^{ -i
\Delta_{1}\tau_d}\cos(\Omega_c \tau_d)+\mu_2^2 e^{ -i
\Delta_{2}\tau_d}+\mu_3^2e^{ -i \Delta_{3}\tau_d}\bigr)
\sin^2\theta\\
 C_1(t_2^+)&=& i \frac{\mu_1}{2\mu}\sin 2\theta\Bigl[1+ \frac{1}{\mu^2}
\Bigl(\mu_1^2 e^{- i \Delta_{1}\tau_d} \cos(\Omega_c \tau_d)
+\sum_{k=2}^3\mu_k^2e^{ -i
\Delta_{k}\tau_d}\Bigr)\Bigr]\nonumber\\
&&+i \frac{\mu_1}{\mu^3} \sin \theta
\Bigl[\Bigl(\sum_{k=2}^3\mu_k^2\Bigr)e^{- i \Delta_{1}\tau_d}
\cos(\Omega_c \tau_d ) -\Bigl(\sum_{k=2}^3\mu_k^2e^{
-i \Delta_{k}\tau_d}\Bigr)\Bigr]\\
C_2(t_2^+) &=&  i \frac{\mu_2}{2\mu} \sin
2\theta\Bigl[1+\frac{1}{\mu^2}\Bigl(\mu_1^2 e^{- i
\Delta_{1}\tau_d}  \cos(\Omega_c \tau_d) +\sum_{k=2}^3\mu_k^2e^{
-i \Delta_{k}\tau_d}\Bigr)\Bigr]
\nonumber\\
&&+i \frac{\mu_2}{\mu^3} \sin \theta \Bigr\{\mu_1^2\Bigl[e^{ -i
\Delta_{2}\tau_d}- e^{- i \Delta_{1}\tau_d} \cos(\Omega_c
\tau_d)\Bigr]+\mu_3^2\bigl(e^{ -i \Delta_{2}\tau_d}-e^{ -i
\Delta_{3}\tau_d}\bigr)\Bigr\}
\\
 C_3(t_2^+) &=& C_{2 \leftrightarrow 3}(t_2^+),\\
C_4(t_2^+)&=&-\frac{\mu_1}{\mu}e^{- i\Delta_{1}\tau_d} \sin \theta
\sin (\Omega_c\tau_d)
\end{eqnarray}
\end{subequations}
\end{widetext}
with
\begin{equation}
\theta=\frac{\mu}{\hbar}\int \mathcal E (t)dt,\quad
\mu=\biggl(\sum_{k=1}^3\mu_k^2\biggr)^{1/2},
\end{equation}
and $\int \mathcal E (t)dt$ the area of each pump pulse
(considered invariant from pulse to pulse). When the upper-lying
states 1,2,3 are harmonic such that (i) $\omega_{ij} \tau_d = 2\pi
n$, (ii) condition (\ref{1}) is fulfilled, and (iii) the pump
pulses are resonant with one of any transitions $0\rightarrow i$,
i.e. $\Delta_i = 0$, and implying $\Delta_j\tau_d=2\pi n_j$ for
all $j$ (with $n_j$ an integer) from condition (i), these
equations take the simpler form
\begin{subequations}
\begin{eqnarray}
C_0(t_2^+)&=&\cos^2 \theta - \frac{\mu_2^2+\mu_3^2 -\mu_1^2}{\mu^2} \sin^2\theta, \\
C_1(t_2^+)&=&2i \frac{\mu_1}{\mu} \frac{\mu_2^2+\mu_3^2}{\mu^2} \sin \theta (\cos \theta - 1),\\
C_2(t_2^+)&=&i \frac{\mu_2}{\mu}\bigl ( \frac{\mu_2^2 + \mu_3^2}{\mu^2}\sin 2 \theta +\frac{2 \mu_1^2}{\mu^2}\sin \theta\bigr),\\
 C_3(t_2^+)& =& C_2, (2 \leftrightarrow 3), C_4(t_2^+) = 0.
\end{eqnarray}
\end{subequations}
This shows that, in order to cancel out the population transfer to
state 1 while increasing the population of states 2 and 3, only
the limit of weak pump excitation ($\theta\ll1$) is suitable since
it leads to $\cos \theta - 1= O(\theta^2)$.

\subsection{Solution in the perturbative regime}

If we consider that each pump pulse is weak: $\Omega_jT\ll1$, we
can perturbatively calculate the solution of Eqs. (\ref{ampl})
(without invoking explicitly the shortness of the pump pulse). We
obtain with correction of order $O(\theta^2)$:
\begin{subequations}
\label{Pert}
\begin{eqnarray}
%\label{itC0_}
%C_0(t_{n+1}^+) &=&  C_0(t_n^+)+\frac{i}{\mu}\theta
%\Bigl\{\mu_1^{\ast}e^{-
%i\Delta_{1}\tau_d}\nonumber\\
%&&\times\Bigl[C_1(t_n^+)\cos(\Omega_c
%\tau_d)+iC_4(t_n^+) \sin (\Omega_c\tau_d)\Bigr]\nonumber\\
%&&+\sum_{k=2}^3\mu_k^{\ast}e^{
%-i \Delta_{k}\tau_d}C_k(t_n^+)\Bigr\},\quad\\
\label{itC1_} C_1(t_{n+1}^+) &=&i\theta_1 +e^{-
i\Delta_{1}\tau_d}\Bigl[C_1(t_n^+)\cos(\Omega_c
\tau_d)\nonumber\\
&&+iC_4(t_n^+) \sin (\Omega_c\tau_d)\Bigr],\\
\label{itCj_} C_j(t_{n+1}^+) &=&i\theta_j +e^{-
i\Delta_{j}\tau_d}C_j(t_n^+),\quad j=2,3\\
\label{itC4_}C_4(t_{n+1}^+)&=&e^{-
i\Delta_{1}\tau_d}\bigl[iC_1(t_n^+) \sin
(\Omega_c\tau_d)\nonumber\\
&&+C_4(t_n^+) \cos (\Omega_c \tau_d)\bigr]
\end{eqnarray}
\end{subequations}
with
\begin{equation}
\theta_j=\frac{\mu_j}{\hbar}\int{\cal E}(t)e^{i\Delta_jt}dt
\end{equation}
 the
Fourier spectral component of the Rabi frequency of the pump pulse
at frequency $\Delta_j$. We remark that we recover these equations
\eqref{Pert} from Eqs. \eqref{itC} using
$\sin\theta=\theta+O(\theta^3)$ and $\cos\theta=1+O(\theta^2)$
except for the phase in the $\theta_j$'s that are neglected in the
impulsive regime.

\subsubsection{Selective excitation to a single state \cite{araujo}}

To excite a single state, say state 1, no control field is
required : $\Omega_c = 0$, and the pump needs to be resonant with
the target state: $\Delta_1=0$. In that case one can determine the
coefficients after $n$ pulses from Eqs. \eqref{Pert}:
\begin{subequations}
\begin{eqnarray}
\label{itC1_se1} C_1(t_{n}^+) &=&in\theta_1,\\
\label{itCj_se1} C_{j=2,3}(t_{n}^+) &=&ie^{-
i(n-1)\Delta_{j}\tau_d/2}\frac{\sin(\frac{n}{2}\Delta_j\tau_d)}
{\sin(\frac{1}{2}\Delta_j\tau_d)}\theta_j . \qquad
\end{eqnarray}
\end{subequations}
This shows that population in the target state accumulates
linearly as a function of the number of the ultrashort pulses.
Population does not coherently accumulate for large $n$ in the
other state if one chooses \textit{$\Delta_j\tau_d$ well different
from $2\pi k$}, $k$ an integer. This effect is optimal when
\begin{equation} \Delta_j\tau_d=\pi(1+2k).
\end{equation}
 The population
transfer to state 1 is closer to 1 when the total area of the pump
pulses is 2$\pi$. (This value obtained here 2$\pi$ is due to the
definition of the fields \eqref{Field} and the Rabi frequencies
\eqref{Rabi}; This corresponds to a ``$\pi$-pulse'' transfer of a
single strong field.) The resulting selective excitation is thus
very robust with respect to $\Delta_j\tau_d$.

\subsubsection{Selective excitation to a superposition of states}

To excite a superposition of states, one has to impose
\begin{equation}
\Delta_j\tau_d=2\pi k_j
\end{equation}
 with $k_j$ an integer, which leads to
\begin{equation}
\label{itCj_se} C_{j=2,3}(t_{n}^+) =ie^{- i(n-1)\pi}n \theta_j .
\end{equation}
This condition can be satisfied when the upper-lying states within
the bandwidth of a single pulse are harmonic.

\begin{figure}[b] \rotatebox{0}{\includegraphics* [scale =
0.9]{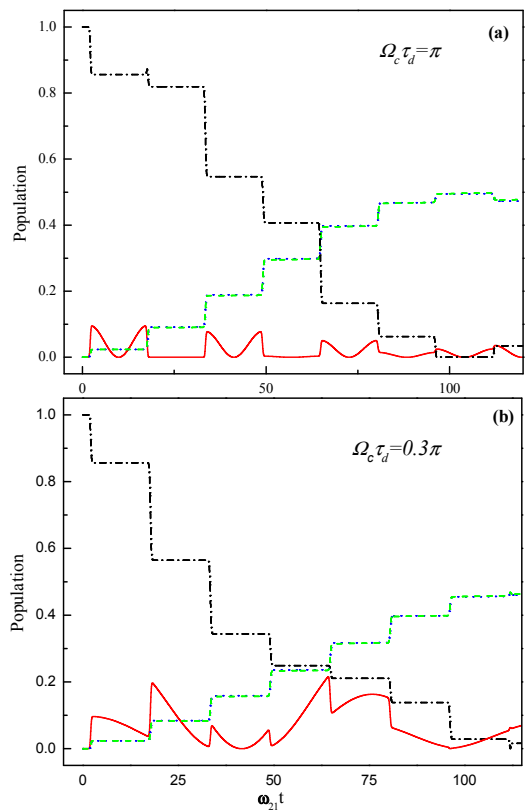}} \caption{(color online)Populations of atomic
ground state (dash-dotted, black) and upper levels 1 (solid, red),
2 (dotted, blue) and 3 (dashed, green) excited by a train of the
pump pulses for $\mu_2=\mu_3=0.5\mu_1, T=0.3\omega_{21}$  and a) $\Omega_{c} \tau_d=\pi$;
b) $\Omega_{c}\tau_d=0.3\pi$.}
\end{figure}

We now show that the control field allows to remove the transition
to the state to which this control field is resonantly coupled. We
choose state 1 to have this feature, i.e. $\Delta_1=0$.  From Eqs.
\eqref{Pert}, we get after $n$ pulses:
\begin{subequations}
\begin{eqnarray}
\label{itC1_se} C_1(t_{n}^+)
&=&i\cos\Bigl(\frac{n-1}{2}\Omega_c\tau_d\Bigr)
\frac{\sin(\frac{n}{2}\Omega_c\tau_d)}{\sin(\frac{1}{2}\Omega_c\tau_d)}\theta_1,\\
\label{itC4_se} C_{4}(t_{n}^+)
&=&i\sin\Bigl(\frac{n-1}{2}\Omega_c\tau_d\Bigr)
\frac{\sin(\frac{n}{2}\Omega_c\tau_d)}{\sin(\frac{1}{2}\Omega_c\tau_d)}\theta_1.
\qquad
\end{eqnarray}
\end{subequations}
The populations do not accumulate in states 1 and 4 if one chooses
\textit{$\Omega_c\tau_d$ well different from $2\pi k$}, $k$ an
integer. This effect is optimal when $\Omega_c\tau_d =\pi(1+2k)$
[see condition \eqref{1}]. The value for $k=0$ corresponds to a
$\pi$ area, i.e. a ``$\pi/2$-pulse'', for the control field in
this model. We remark that such a cancelation
of the transfer to state 1 is thus expected to be robust with
respect to a precise area of the control field.

Thus, by choosing the number of the pump-pulses, one can achieve
the coherent selective superposition of the levels 2 and 3, while
keeping the state 1 almost empty.

In Fig. 2a we show the results of numerical integration of Eqs.
(\ref{ampl}) obtained under the conditions mentioned above using
Gaussian shape $f(t)=\exp(-t^2/T^2)$ for the pump pulses. Very
similar results are obtained, when the condition (\ref{1}) is
significantly violated, as shown in Fig. 2b. This demonstrates the
robustness of our scheme with respect to the coupling field
instabilities as predicted above.

\begin{figure}[b] \rotatebox{0}{\includegraphics* [scale =
0.9]{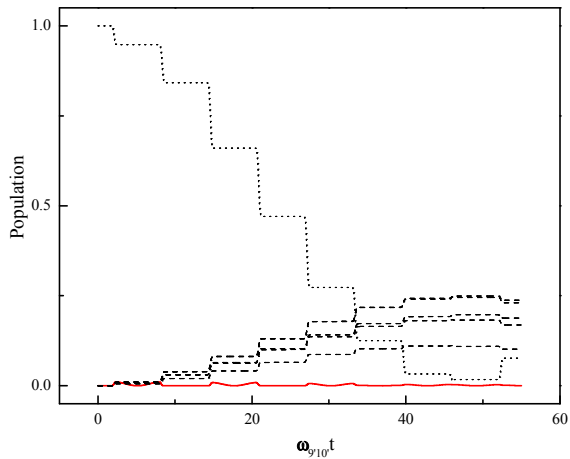}} \caption{(color online) Populations of $K_2$ of
the ground state $v''=0$ (dotted, black), the upper state $v'=10$
(solid, red) and the other states $v'=8,9,11,12,13$ (dashed
lines).}
\end{figure}

We apply the proposed mechanism in the next section to produce a
selective coherent superposition of vibrational states in a
molecular electronic state.

\section{\protect\normalsize Application to the potassium dimer}

We consider the excitation of the potassium dimer $K_2$
\cite{lyyra}. The molecule is supposed to be prepared in the
ground vibrational state $v''=0$ of the electronic state
$X^1\Sigma_g^+$. The excited state is chosen to be the first
excited electronic state $A^1\Sigma_u^+$ (of lifetime 28ns). For
simplicity in the calculations the dependence of the electric
dipole moment on internuclear distance is ignored. The pump pulses
are assumed to be transform limited of Gaussian envelope $f(t) =
\exp(-t^2/T^2)$ with duration $T = 150$fs and peak intensity
$I_{p}^{\max}\sim 10^{11}$ W/cm$^2$. We assume the pump pulses to
be on resonance with the transition $v''=0 \rightarrow v'=10$
($\omega_L \simeq 11 800 cm^{-1}$). The excited vibrational levels
$v'=8, 9, ..., 13$  are within the spectrum of the pump field and
are expected to be populated. Our main goal is to suppress the
strongest transition $v'=10$ of the upper vibrational levels.
\begin{figure}[b] \rotatebox{0}{\includegraphics* [scale =
0.8]{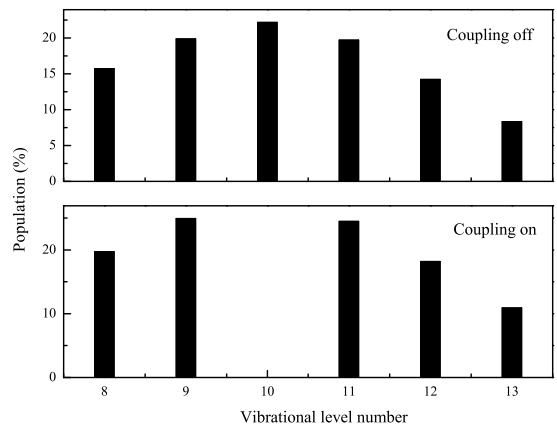}} \caption{Histogram of the vibrational population
distribution after excitation  without (upper frame) and with
(lower frame) the coupling field. }
\end{figure}
\begin{figure}[t] \rotatebox{0}{\includegraphics* [scale =
0.8]{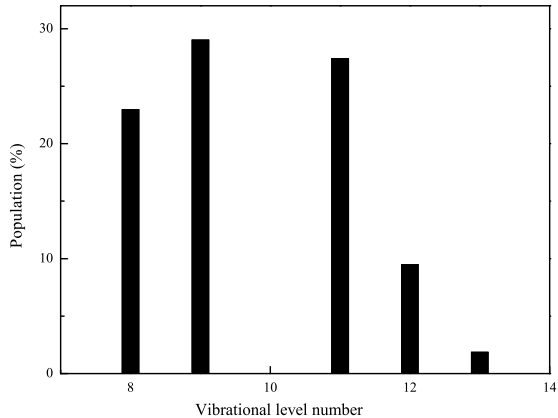}} \caption{Histogram of the vibrational population
distribution after excitation by the same pulses used in Fig. 4
but of duration 600 fs.}
\end{figure}
\begin{figure}[t] \rotatebox{0}{\includegraphics* [scale =
0.8]{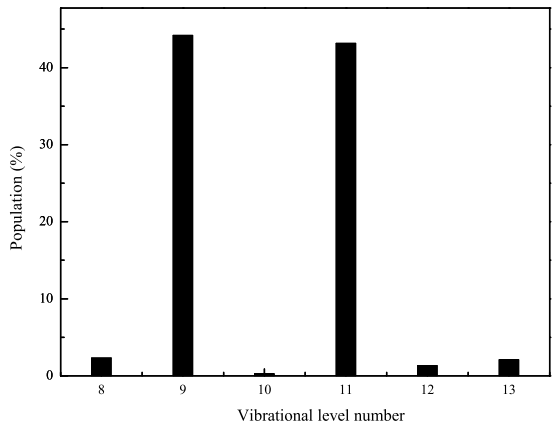}} \caption{Histogram of the vibrational population
distribution when the states $v'=8,10,12,13$ are coupled to
auxiliary states by coupling fields.}
\end{figure}

We solve numerically equations for atomic population amplitudes similar to Eq. \eqref{ampl},
including all the relevant vibrational states of the problem,
using the above parameters, and with the requirement that the
conditions (\ref{1}) and $\omega_{9'10'} = 2\pi k \tau_d$ are
fulfilled, where $\omega_{9'10'}$ is the frequency splitting of
the vibrational states $v'=9$ and $v'=10$ of the upper electronic
state. For $\omega_{9'10'}=67.3 \text{cm}^{-1}$ the delay time between the subpulses is $\tau_d \simeq 3 \text {ps}$. Note, that frequency splitting of the upper lying levels is almost equidistant. The coupling field couples the state $v'=10$, of largest
dipole moment element among the states within the bandwidth of a
single pump, with an auxiliary electronic state of the potassium
dimer, e.g. $b^3\Pi_u$. Figure 3 shows the dynamics of the
populations of the vibrational levels, when it is excited by a
train of identical pulses. To calculate the populations we have used the Franck-Condon factors and the corresponding eigenfrequencies, which are well known for vibrational levels of $K_2$ molecules \cite{lyyra}. The chosen values of the parameters:
$\Omega_c\simeq 0.2 \omega_{9'10'}\ll \Omega_{p}^{max}\simeq 0.6
\omega_{9'10'}$ and $\theta^2(\infty)\sim 0.15$ provide all the
necessary conditions for the analytical analysis made in the
previous section to be valid. As it is seen in Fig. 3, after the
interaction of the molecule with 8 pulses, all the population is
distributed between the upper vibrational levels $v'=8,9,11,12,13$
(dotted lines), while the level $v'=10$ (red, solid line) stays
almost unpopulated. To have a more complete picture of the
process, Fig. 4 displays the histogram of the population
distribution in the upper-lying states. In the absence of the
coupling field the population is distributed between all upper
vibrational states (Fig. 4 upper frame). But if the coupling field
is on (Fig. 4 lower frame), the strongest transition is
dramatically suppressed. Adapting the duration of the pulses
allows one to modify the shape of the superposition, reducing the
population of the upper states, as it is shown in Fig. 5. Here the
duration of the pump pulses are taken to be four times larger than
that of the cases considered previously. To suppress some
components of the superposition one can apply other fields coupled
to the undesired states from different auxiliary states. As an
example in Fig. 6 the levels $v'=8,10,12,13$ are coupled with
other molecular states which leads to a coherent superposition of
only two states $v'=9,11$.
Thus, we have shown that the coupling fields allow the decreasing
of the populations of the undesired states and the enhancement of
the populations of the other states well within the bandwidth of a
single pump.

\section{\protect\normalsize Conclusion}

In this paper we have proposed a robust and simple mechanism for
the coherent excitation of the molecule or atom to a superposition
of pre-selected states by a train of fs laser pulses, combined
with narrow-band weak laser fields coupling the undesired states
well within the bandwidth of a single pulse to auxiliary states.
The coupling fields allow the cancelation of specific transitions
from the ground state to a set of states $i$ when they induce a
coherence between each state $i$ and an auxiliary state (all
different). We remark that these predictions of selective coherent
excitation could be measured experimentally by a sensitive method
such as the one developed in \cite{gogyan}.

\bigskip
\subsection*{Acknowledgments}%\nonumber

This research has been conducted in the scope of the International Association Laboratory IRMAS. We also acknowledge the support from the Armenian Science Ministry (Grant No 096), the French Agence Nationale de la Recherche (Project CoMoC) and the Marie Curie Initial Training Network Grant No. CA-ITN-214962-FASTQUAST.

\bigskip
\appendix*
\section{\protect \normalsize General solution in the impulsive regime}

%\section{Appendix: General solution in the impulsive regime}

We integrate Eqs. (\ref{ampl}) over the time of interaction with
the n$^{\text{th}}$ pump pulse in the impulsive approximation,
disregarding the detunings $\Delta_jT\ll1$ and where to a good
approximation the weak $\Omega_c$ - terms can be neglected. This
yields a simple solution right after the n$^{\text{th}}$ pump
pulse at time $t=t_n^+$ (considered interacting at $t=t_n$), from
the solution right before the pulse at time $t=t_n^-$:
\begin{subequations}
\begin{eqnarray}\label{am0reg1}
C_0(t_n^+) &=&  C_0(t_n^-)\cos \theta+i\bar C_n^-\sin\theta\\
\label{am3reg1} C_j(t_n^+) &=& C_j(t_n^-)+ i\frac{\mu_j}{\mu}
C_0(t_n^-)
\sin \theta\nonumber\\
+&& \frac{\mu_j}{\mu}\bar C_n^- (\cos \theta-1),\quad j=1,2,3\\
\label{am4reg1}C_4(t_n^+)&=&C_4(t_n^-),
\end{eqnarray}
\end{subequations}
where $\mu=\sqrt{\sum_{k=1}^3\mu_k^2}$,
$\theta=\frac{\mu}{\hbar}\int \mathcal E (t)dt$ with $\int
\mathcal E (t)dt$ the area of each pump pulse (considered
invariant from pulse to pulse), and $\bar
C_n^-=[\sum_{k=1}^3\mu_k C_k(t_n^-)]/\mu$. After the
n$^{\text{th}}$ pulse turned off, the amplitudes $C_0$, $C_2$ and
$C_3$ evolve freely up to $t \sim t_n+\tau_d$, $\tau_d\gg T$:
\begin{subequations}
\begin{eqnarray}
C_0(t) &=& C_0(t_n^+) ,\\
C_j(t) &=& e^{ -i \Delta_{j} (t-t_n)}C_j(t_n^+),\ j=2,3,
\end{eqnarray}
 while $C_1(t)$ and $C_4(t)$, $t_{n+1}>t>t_n$ are
found from Eqs. (\ref{ampl}) with the initial values
(\ref{am3reg1}-\ref{am4reg1}) as
\begin{eqnarray}
\label{am1reg2} C_1(t) &=& e^{- i \Delta_{1}
(t-t_n)}\Bigl\{C_1(t_n^+) \cos \bigl [ \Omega_c (t-t_n) \bigr ]\nonumber\\
&&+iC_4(t_n^+) \sin \bigl [ \Omega_c (t-t_n) \bigr ]\Bigr\},\\
\label{am3reg2} C_4(t) &=& e^{ -i \Delta_{1} (t-t_n)}
\Bigl\{iC_1(t_n^+)
\sin \bigl [\Omega_c (t-t_n) \bigr ]\nonumber\\
&&+C_4(t_n^+) \cos \bigl [ \Omega_c (t-t_n) \bigr ]\Bigr\},
\end{eqnarray}
\end{subequations}
 We iterate the
above procedure for all the ultrashort pump pulses:
\begin{widetext}
\begin{subequations}
\label{itC}
\begin{eqnarray}\label{itC0}
C_0(t_{n+1}^+) &=&  C_0(t_n^+)\cos \theta+\frac{i}{\mu}\sin\theta
\Bigl\{\mu_1e^{-
i\Delta_{1}\tau_d}\Bigl[C_1(t_n^+)\cos(\Omega_c
\tau_d)+iC_4(t_n^+) \sin (\Omega_c\tau_d)\Bigr]
+\sum_{k=2}^3\mu_ke^{
-i \Delta_{k}\tau_d}C_k(t_n^+)\Bigr\},\qquad\ \ \\
\label{itC1} C_1(t_{n+1}^+)
&=&i\frac{\mu_1}{\mu}C_0(t_n^+)\sin\theta +e^{-
i\Delta_{1}\tau_d}\Bigl[C_1(t_n^+)\cos(\Omega_c
\tau_d)+iC_4(t_n^+) \sin (\Omega_c\tau_d)\Bigr]
\Bigl[1+\frac{\mu_1^2}{\mu^2}(\cos\theta-1)\Bigr]\nonumber\\
&&+\frac{\mu_1}{\mu}(\cos
\theta-1)\sum_{k=2}^3\frac{\mu_k}{\mu}e^{
-i \Delta_{k}\tau_d}C_k(t_n^+),\\
\label{itC2} C_2(t_{n+1}^+)
&=&i\frac{\mu_2}{\mu}C_0(t_n^+)\sin\theta +e^{-
i\Delta_{2}\tau_d}C_2(t_n^+)\Bigl[1+\frac{\mu_2^2}{\mu^2}(\cos\theta-1)\Bigr]\nonumber\\
&&+\frac{\mu_2}{\mu}(\cos
\theta-1)\Bigl\{\frac{\mu_1}{\mu}e^{ -i
\Delta_{1}\tau_d}\Bigl[C_1(t_n^+)\cos(\Omega_c \tau_d)+iC_4(t_n^+)
\sin (\Omega_c\tau_d)\Bigr]+\frac{\mu_3}{\mu}e^{ -i
\Delta_{3}\tau_d}C_3(t_n^+)\Bigr\},\\
\label{itC3} C_3(t_{n+1}^+) &=& C_{2 \leftrightarrow 3}(t_{n+1}^+),\\
\label{itC4}C_4(t_{n+1}^+)&=&e^{-
i\Delta_{1}\tau_d}\bigl[iC_1(t_n^+) \sin
(\Omega_c\tau_d)+C_4(t_n^+) \cos (\Omega_c \tau_d)\bigr].
\end{eqnarray}
\end{subequations}
\end{widetext}
Eq. (\ref{itC3}) means that the amplitude $C_3$ has the same
expression as $C_2$ \eqref{itC2} but exchanging the indices 2 and
3.

\end{document}